\documentclass[aps,prd,twocolumn,preprintnumbers,showpacs,superscriptaddress,nofootinbib,floatfix,10pt]{revtex4-1}
\pdfoutput=1
\usepackage{textcomp}
\usepackage[english]{babel}
\usepackage{amsmath,amssymb,amsbsy,booktabs}
\usepackage{bm}
\usepackage{color}
\usepackage{array}
\usepackage{amstext}
\usepackage{graphicx}
\usepackage{amsfonts}
\usepackage{bm}
\usepackage{dcolumn}
\usepackage{rotating}
\usepackage{epstopdf}
\usepackage{esint}
\usepackage{url}
\usepackage[colorlinks=true,
            linkcolor=blue,
            filecolor=blue,
            anchorcolor=blue,
            urlcolor=blue,
            citecolor=blue
            ]{hyperref}

\usepackage[utf8]{inputenc}
\usepackage{float}
\usepackage{subfigure}

\usepackage[T1]{fontenc} % Output font encoding for international characters

\allowdisplaybreaks[0]

\begin{document}

\title {\bf Simple estimate of BBN sensitivity to light freeze-in dark matter}

\author{Shao-Ping Li}
\email{ShowpingLee@mails.ccnu.edu.cn}
\affiliation{Institute of Particle Physics and Key Laboratory of Quark and Lepton Physics~(MOE),\\
	Central China Normal University, Wuhan, Hubei 430079, China}

\author{Xin-Qiang Li}
\email{xqli@mail.ccnu.edu.cn}
\affiliation{Institute of Particle Physics and Key Laboratory of Quark and Lepton Physics~(MOE),\\
	Central China Normal University, Wuhan, Hubei 430079, China}

\author{Xin-Shuai Yan}
\email{xinshuai@mail.ccnu.edu.cn}

\author{Ya-Dong Yang}
\email{yangyd@mail.ccnu.edu.cn}
\affiliation{Institute of Particle Physics and Key Laboratory of Quark and Lepton Physics~(MOE),\\
	Central China Normal University, Wuhan, Hubei 430079, China}

\begin{abstract}
We provide a simple analysis of the big-bang nucleosynthesis (BBN) sensitivity to the light dark matter (DM) generated by the thermal freeze-in mechanism. It is shown that the ratio of the effective neutrino number shift $\Delta N_{\nu}$ over the DM relic density $\omega\equiv \Omega h^2$, denoted by $R_\chi\equiv\Delta N_\nu/\omega$, cancels the decaying particle mass and the feeble coupling, rendering therefore a simple visualization of $\Delta N_{\nu}$ at the BBN epoch in terms of the DM mass. This property drives one to conclude that the shift with a sensitivity of $\Delta N_{\nu}\simeq \mathcal{O}(0.1)$ cannot originate from a single warm DM under the Lyman-$\alpha$ forest constraints. For the cold-plus-warm DM scenarios where the Lyman-$\alpha$ constraints are diluted, the ratio $R_\chi$ can be potentially used to test the thermal freeze-in mechanism in generating a small warm component of DM and a possible excess at the level of $\Delta N_{\nu}\simeq \mathcal{O}(0.01)$.

\end{abstract}


\maketitle

\section{Introduction} 

Hidden light particles with typical masses around keV scale are widely considered as the dark matter (DM) candidates~\cite{Adhikari:2016bei}. Such light particles are usually associated with feeble couplings and have a non-thermal history in the early Universe. Feeble interactions with the Standard Model (SM) sector are extremely difficult to probe directly at colliders and indirectly in flavor physics. Nevertheless, the precision era of cosmology now provides us with striking opportunities to search indirectly for these hidden particles. Among other things, the precise measurements of light elements produced at the big-bang nucleosynthesis (BBN) epoch set constraints on any new physics effects that can cause a significant shift of the effective neutrino number $\Delta N_\nu\equiv N_{\nu}-N_\nu^{\rm SM}$ away from the SM prediction $N_\nu^{\rm SM}=3$. When the BBN measurements, including the primordial abundances of helium-4 and deuterium, are combined with the cosmic microwave background (CMB) data (CMB+BBN+$Y_p$+D), the constraints become even tighter and $N_{\nu}$ is found to be~\cite{Fields:2019pfx}:
\begin{align}
  N_{\nu}=2.843\pm 0.154,
\end{align} 
which sets a $2\sigma$ upper bound on the shift $\Delta N_\nu<0.151$.

The generation of an observable shift of $\Delta N_{\nu}$ at the BBN epoch or of $\Delta N_{eff}$ at the CMB regime with respect to the SM prediction $N_{eff}^{\rm SM}=3.045$ obtained after taking into account the non-instantaneous neutrino decoupling effects~\cite{Mangano:2005cc,deSalas:2016ztq,Escudero:2020dfa,Akita:2020szl,Froustey:2020mcq,Bennett:2020zkv}, has opened up an avenue to probe light hidden species in the early Universe (see, e.g., Refs.~\cite{Hooper:2011aj,Roland:2016gli,Shakya:2016oxf,Sabti:2019mhn,Berlin:2019pbq,Abazajian:2019oqj,Luo:2020sho,Luo:2020fdt}). In particular, for the DM scenarios with a non-thermal history, such as the representative freeze-in evolution~\cite{Hall:2009bx,Bernal:2017kxu}, if the  hidden species reside at keV or sub-keV scale, they could contribute to the Hubble expansion as extra radiation prior to the BBN epoch and become non-relativistic before the recombination. Therefore, the BBN measurements can provide a promising avenue to probe and/or constrain the non-thermal light DM which nevertheless has a feeble connection to the visible world.

Here, we present a simple analysis of the BBN sensitivity to the non-thermal light species which is generated by the thermal freeze-in mechanism. While the BBN constraints on non-thermal keV warm DM have been partially discussed by specifying the DM momentum distribution~\cite{Merle:2015oja,Ballesteros:2020adh}, we will not refrain ourselves from a single warm DM candidate and extend the analysis to any light hidden species that are difficult to probe directly at terrestrial colliders and indirectly in flavor physics. Noticeably, our analysis could be powerful in testing the mixed cold-plus-warm DM scenarios (see e.g., Refs.~\cite{Palazzo:2007gz,Boyarsky:2008xj,Diamanti:2017xfo,Murgia:2017lwo,Gariazzo:2017pzb}) in which the small-scale problems encountered in cold DM simulations can be alleviated~\cite{Weinberg:2013aya}. As a simple application in this context, our analysis can be used as an economic criterion to test whether a small warm DM component and an observable $\Delta N_\nu$ excess have the common source from a sub-keV hidden species. The analysis presented here is based on the integrated Boltzmann equations of the energy and the particle-number density evolution, instead of solving the unintegrated kinetic equations of momentum distribution. 

As will be shown later, the current observation of the DM relic density allows us to see directly how large a shift of $\Delta N_{\nu}$ can be generated. Remarkably, this visualization is basically independent of the decaying particle mass and the feeble coupling, but essentially dependent on the relic DM mass. Such a \textit{freeze-in independence} property can promote a simple avenue to test the thermal freeze-in mechanism as the common source of $\Delta N_{\nu}$ excess and the DM relic density. Given the severe bounds from the Lyman-$\alpha$ forest data (see e.g., Refs.~\cite{Narayanan:2000tp,Viel:2005qj,Viel:2013fqw,Baur:2015jsy,Irsic:2017ixq,Palanque-Delabrouille:2019iyz}), we can readily conclude that a single warm DM cannot generate an observable shift of $\Delta N_{\nu}$ under the current and future sensitivities. Nevertheless, since the Lyman-$\alpha$ constraints are diluted in the mixed cold-plus-warm DM scenarios~\cite{Boyarsky:2008xj,Diamanti:2017xfo,Gariazzo:2017pzb}, we find that the forecast $\Delta N_\nu \sim \mathcal{O}(0.01)$ from a combination of BBN and CMB-S4~\cite{Abazajian:2016yjj,Abazajian:2019eic} is able to test a sub-keV DM that makes up only a small fraction of the current relic density.

It should be noted that our analysis presented here could also be generalized to other DM production mechanisms, though it may not be always true to obtain the simple \textit{freeze-in independence}. Some discussions about the validity of such a \textit{freeze-in independence} will be presented before the conclusion section.

\section{Effective neutrino number shift from extra radiation}  

The primary epoch of BBN takes place when the temperature of the early Universe cools down to a few MeV and lasts till $\mathcal{O}(10)$~keV (see Ref.~\cite{Cyburt:2015mya} for a review). Any extra radiation contributing to the SM plasma at this epoch can be parameterized by the variation of effective neutrino number: 
\begin{align}\label{Nnu-def}
  \Delta N_{\nu}\equiv \frac{\rho_{rad}}{\rho_{\nu_i}},
\end{align}
which is normalized to the energy density of one active neutrino flavor $\rho_{\nu_i}$. The main effect of $\Delta N_{\nu}$ is to modify the Hubble rate through the total energy density, which further affects the freezing temperature of neutron. As a consequence, the neutron-to-proton ratio is altered and a different abundance of helium-4 from what is predicted by the SM with $N_\nu^{\rm SM}=3$ would be generated.
  
In the SM where $N_\nu$ is fixed, the mass fraction of helium-4, denoted by $Y_p$, depends on the baryon density ($\eta_b$) and the neutron mean lifetime ($\tau_n$). When $N_\nu$ varies due to the extra radiation contribution, one can use the BBN code to obtain a semi-analytic formula of $Y_p(N_\nu)$ by fixing the two free parameters $\eta_b$ and $\tau_n$ to their experimental measurements~\cite{Sarkar:1995dd,Cyburt:2015mya,Fields:2019pfx}. Then, a bound on $\Delta N_\nu$ can be extracted by comparing   $Y_p(N_\nu)$ with the measured abundance. With this method, it has been shown that the variation of the neutrino number is currently bounded to be $\Delta N_\nu=N_\nu-3<0.151$ at $2\sigma$ level from CMB+BBN+$Y_p$+D~\cite{Fields:2019pfx}. However, it should be mentioned that the direct application of this limit to Eq.~\eqref{Nnu-def} is valid only for the extra radiation that decouples from the SM plasma before the BBN epoch. For the models, e.g., in which the hidden sector thermalizes with photons, electrons or neutrinos during the BBN process (see, e.g., Refs.~\cite{Boehm:2013jpa,Berlin:2019pbq,Sabti:2019mhn}), a simple parameterization of $\Delta N_{\nu}$ is still absent~\cite{Kolb:1986nf}. In these situations, one should perform a fully numerical simulation via the BBN code (see, e.g., Refs.~\cite{Wagoner:1966pv,Pisanti:2007hk,Consiglio:2017pot,Arbey:2011nf,Cyburt:2015mya}). For the non-thermal light species considered  here, we will assume that they decouple already from the SM plasma before the BBN begins. This allows us to find a simple parametrization of $\Delta N_{\nu}$ and probe the corresponding signals and/or constraints of the hidden sector. Note that we do not consider the shift of $\Delta N_{eff}$ at the CMB epoch, since we are interested in the light DM that is relativistic at the BBN epoch but becomes non-relativistic prior to the recombination. 

In practice, we can express $\rho_{\nu_i}$ in Eq.~\eqref{Nnu-def} in terms of the total relativistic energy density of the SM plasma $\rho_{SM}$:
\begin{align}
 \rho_{SM}=g_*^\rho \frac{\pi^2}{30}T^4,
\end{align} 
where $g_*^\rho$ denotes the effective number of relativistic degrees of freedom (d.o.f) for the energy density. Using the fact that the SM plasma conserves entropy after the hidden sector decouples, we can determine the shift $\Delta N_{\nu}$ in terms of the SM and hidden energy densities:
\begin{align}\label{Neff def-2}
  \Delta N_{\nu}=\frac{4}{7} \left[\frac{g_{*}^s(T_{\gamma e \nu})}{g_{*}^s(T_{ref})}\right]^{4/3} g_*^\rho (T_{ref})\, \frac{\rho_{rad}(T_{ref})}{\rho_{SM}(T_{ref})},
\end{align}
where $T_{ref}$ denotes any reference temperature after the hidden radiation freezes in at $T_{fi}$ and before the species transits to non-relativistic matter at $T_{nr}$. Here the effective d.o.f for the SM entropy density, $g_{*}^s(T_{\gamma e \nu})=10.75$, corresponds to the epoch when the relativistic SM plasma contains photons, electrons, positrons and neutrinos. Using Eq.~\eqref{Neff def-2}, we can then probe and/or constrain new physics effects encoded in $\rho_{rad}$, by comparing the resulting shift $\Delta N_{\nu}$ with the bound extracted from BBN precision measurements.

\section{Particle-number and energy density evolution}

To see how the freeze-in hidden light species contributes to the shift $\Delta N_{\nu}$, we proceed to calculate the particle-number and energy density evolution in light of the integrated Boltzmann equations. For definiteness, we assume that the production is via a thermal scalar decay into two chiral fermions, $\phi\to \chi+\psi$, from the interaction Lagrangian given by
\begin{align}
  \mathcal{L}=-y\,\bar \psi_L\phi\, \chi_R + \rm H.c.,
\end{align} 
where $y\ll 1$ is a feeble Yukawa-like coupling. For simplicity, the fermion $\chi$ is assumed to be the DM candidate while $\phi$ and $\psi$ are non-DM species. 

Let us now concentrate on the $\chi$ evolution. It is, as usual, convenient to simplify the Boltzmann equation via the particle yield $Y_{\chi,n}\equiv n_\chi/s$, with the SM entropy density given by
\begin{align}
  s=g_*^s\,\frac{2\pi^2}{45}\,T^3.
\end{align}
The reduced Boltzmann equation for the $\chi$ species reads
\begin{align}\label{n-Beq}
  \frac{dY_{\chi,n}}{dT}=-\frac{C_{\chi,n}}{sH T},
\end{align}
where the Hubble rate is given by $H\approx1.66\sqrt{g_*^\rho}\,T^2/M_{\mathrm{Pl}}$, with the Planck mass $M_{\mathrm{Pl}}\approx1.22\times 10^{19}$~GeV. The collision term $C_{\chi,n}$ is calculated in the center-of-mass frame of the thermal scalar $\phi$ with internal d.o.f $g_\phi=1$, and reads
\begin{align}\label{C-n}
	C_{\chi,n}&\approx\int \frac{d^3p_\phi}{(2\pi)^32E_\phi}  f_\phi   \int \frac{d^3p_\chi}{(2\pi)^32E_\chi} \frac{d^3p_\psi}{(2\pi)^32E_\psi}
	\nonumber \\[0.15cm]
&\qquad \times (2\pi)^4\,\delta^4(p_\phi-p_\chi-p_\psi)\, \vert \mathcal{M}_{\phi\to \chi \psi} \vert^2
	\nonumber \\[0.15cm]
&\approx\frac{m_\phi^3 y^2}{32\pi^3}\, T K_1(m_\phi/T),
\end{align}
where the Boltzmann distribution $f_\phi=e^{-E_\phi/T}$ has been used, and the inverse decay $\chi+\psi\to \phi$ as well as the Pauli-blocking effects have been neglected, with $1-f_{\chi,\psi}\approx 1$. Here $K_1$ is the first modified Bessel function of the second kind. Integrating Eq.~\eqref{n-Beq} over the temperature $T$ gives the final result of the yield $Y_{\chi,n}$:
\begin{align}\label{Y-n}
	Y_{\chi,n} &\approx\int_0^{\infty}\frac{C_{\chi,n}}{sH T}dT
	\nonumber \\[0.15cm]
	&\approx 7.83\times 10^{16}\, g_{*,\chi dec}^{-3/2}\, y^2\, \left(\frac{\text{GeV}}{m_\phi}\right),
\end{align}
where we have made the approximation $g_*^\rho \approx g_*^s\approx g_{*,\chi dec}$, corresponding to the effective d.o.f at the $\chi$ decoupling temperature $T_{\chi dec}$. The reason behind such an approximation is that, since the yield accumulation is culminated around $T_{\chi dec}\simeq\mathcal{O}(m_\phi)$, for $m_\phi\gg \text{MeV}$ concerned here, the variations of $g_*^\rho$ and $g_*^s$ with respect to the temperature are sufficiently small~\cite{Borsanyi:2016ksw}, and the heavier $m_\phi$ is, the better the approximation will be. In addition, we have fixed the lower limit of the temperature integration to be zero, while a lower limit fixed at the freeze-in temperature $T_{fi}\lesssim m_\phi$ results in a negligible numeric difference. This is a generic feature of the freeze-in mechanism as the production is strongly Boltzmann suppressed when $T\ll m_\phi$.

The evolution of the energy density can be simplified analogously. To this end, we can define the energy yield $Y_{\chi,\rho}\equiv \rho_\chi/s^{4/3}$, so that after $\chi$ decouples relativistically from the SM plasma the yield is scale-independent until $\chi$ becomes non-relativistic. We have then a Boltzmann equation similar to Eq.~\eqref{n-Beq}:
\begin{align}\label{e-Beq}
	\frac{dY_{\chi,\rho}}{dT}=-\frac{C_{\chi,\rho}}{s^{4/3}H T},
\end{align}
where the collision term $C_{\chi,\rho}$ in the center-of-mass frame of $\phi$ is evaluated to be 
\begin{align}\label{C-rho}
	C_{\chi,\rho}
	&\approx\int \frac{d^3p_\phi}{(2\pi)^32E_\phi}  f_\phi   \int \frac{d^3p_\chi}{(2\pi)^32E_\chi} \frac{d^3p_\psi}{(2\pi)^32E_\psi}
	\nonumber \\[0.15cm]
	&\qquad \times	(2\pi)^4\delta^4(p_\phi-p_\chi-p_\psi) E_\chi\vert \mathcal{M}_{\phi\to \chi \psi} \vert^2
	\nonumber \\[0.15cm]
	&\approx \frac{m_\phi^4 y^2}{64\pi^3}\, T K_1(m_\phi/T).
\end{align}
Note that the approximations made in deriving $C_{\chi,n}$ are again applied here. The final result of the energy yield $Y_{\chi,\rho}$ is given by
\begin{align}\label{Y-rho}
	Y_{\chi,\rho}&\approx\int_0^{\infty}\frac{C_{\chi,\rho}}{s^{4/3}H T}dT
	\nonumber \\[0.15cm]
	&\approx 1.75\times  10^{17}\,  g_{*,\chi dec}^{-11/6}\, y^2\, \left(\frac{\text{GeV}}{m_\phi}\right).
\end{align}

It can be immediately seen from Eqs.~\eqref{Y-n} and \eqref{Y-rho} that
\begin{align}\label{Y-ratio}
	\frac{Y_{\chi,\rho}}{Y_{\chi,n}} \approx 2.23\, g_{*,\chi dec}^{-1/3}, 
\end{align}
which is now basically independent of the decaying particle mass $m_\phi$ and the feeble Yukawa-like coupling $y$ (As the decoupling temperature $T_{\chi dec}\simeq \mathcal{O}(m_\phi)$, an indirectly weak dependence is still observed through the effective d.o.f $g_{*,\chi dec}$, which nevertheless causes only a negligible effect for $m_\phi\gg \text{MeV}$ considered here.). Noticeably, we have checked that this \textit{freeze-in independence} is also valid when the Bose-Einstein statistics is used for $\phi$ (but without the model-dependent thermal mass corrections), up to a small numerical change of the prefactor in Eq.~\eqref{Y-ratio}. Thus, we can infer from this property that the ratio $\Delta {N_{\nu}}/{\omega_\chi}$ (to be defined later) would essentially depend only on the DM mass $m_\chi$. As will be shown later, this renders a fast estimate of how large a shift of $\Delta N_\nu$ can be generated by varying the DM mass within the current relic density $\omega_{\rm DM}\equiv \Omega_{\rm DM} h^2=0.120\pm 0.001$~\cite{Aghanim:2018eyx}.

\section{Fast estimate of neutrino number shift} 

Concerning the shift $\Delta N_\nu$ just prior to the BBN epoch, we can apply Eq.~\eqref{Neff def-2} with a reference temperature $T_{ref}=T_{\gamma e\nu}$, leading to 
\begin{align}
	\Delta N_{\nu}=\frac{4}{7}\, \frac{(g_*^s(T_{\gamma e\nu}) 2\pi^2/45)^{4/3}}{\pi^2/30}\, Y_{\chi,\rho}.
\end{align}
The relic density of $\chi$ at present day is estimated to be
\begin{align}\label{relic density}
	\omega_\chi\equiv \Omega_\chi h^2=\frac{Y_{\chi,n}\,s_0\, m_\chi}{\rho_c/h^2}.
\end{align}
Using the current densities $s_0=2891.2~\text{cm}^{-3}$ and $\rho_c= 1.05\times 10^{-5}~h^2\cdot\text{GeV}\cdot \text{cm}^{-3}$~\cite{Zyla:2020zbs} as input, we can determine the ratio $R_\chi\equiv\Delta N_{\nu}/\omega_\chi$, with the final result given by
\begin{align}\label{NOratio}
	R_\chi\equiv \frac{\Delta N_{\nu}}{\omega_\chi}\approx 0.11\, g_{*,\chi dec}^{-1/3}\, \left(\frac{\text{keV}}{m_\chi}\right),
\end{align}
which depends on the DM mass $m_\chi$, and has only a weak dependence on the decaying mass $m_\phi$ through the effective d.o.f $g_{*,\chi dec}$. Equation~\eqref{NOratio} is the main result obtained in this work for the thermal freeze-in two-body decay. Combining with the DM relic abundance, we can then estimate the contribution to $\Delta N_{\nu}$ in light of the hidden light particle mass, but without a sensitive dependence on the decaying mass and the feeble coupling, both of which are directly responsible for the DM production. It should be mentioned that, if the antiparticle $\bar \chi$ is not identical to $\chi$, we should multiply $Y_{\chi,\rho(n)}$ by a factor of two to account for the total energy (number) densities of particle and antiparticle, assuming no significant production of a $\chi$-asymmetry. In addition, if the scalar $\phi$ is a gauge multiplet at the symmetric phase, the collision term $C_{\chi,\rho(n)}$ should be multiplied by the number of gauge components. Nevertheless, the ratio $R_\chi$ defined by Eq.~\eqref{NOratio} remains unchanged in both of these two cases.

\begin{figure}[ht]
	\centering	
	\includegraphics[scale=0.93]{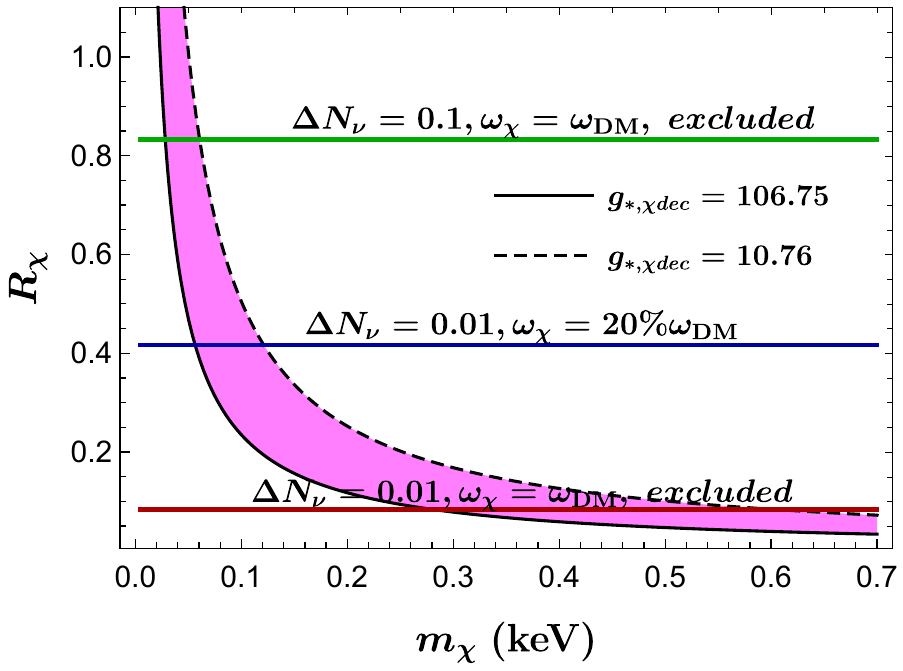}
	\caption{\label{Rmchi-contour} Variation of the ratio $R_\chi$ with respect to the mass of the light hidden species $\chi$, with the magenta band obtained by varying the freeze-in temperature from the electroweak scale ($g_{*,\chi dec}=106.75$) down to $10$~MeV ($g_{*,\chi dec}=10.76$). The current CMB+BBN+$Y_p$+D sensitivity is represented by $\Delta N_{\nu}=0.1$, while the future BBN+CMB-S4 sensitivity by $\Delta N_{\nu}=0.01$. The light species $\chi$ is assumed to make up all (excluded) or, as an example, $20\%$ of the current DM relic density.}
\end{figure}

The current CMB+BBN+$Y_p$+D data sets a limit on $\Delta N_{\nu}$ at the level of $\mathcal{O}(0.1)$~\cite{Fields:2019pfx}, while the future sensitivity from BBN+CMB-S4 is expected to be $\Delta N_{\nu}=\mathcal{O}(0.01)$, depending on the sky fraction $f_{sky}$~\cite{Abazajian:2016yjj,Abazajian:2019eic}. With these sensitivities kept in mind, we show in Fig.~\ref{Rmchi-contour} the variation of the ratio $R_\chi$ with respect to the mass of the light hidden species $\chi$, where the magenta band is obtained by varying the freeze-in temperature from the electroweak scale (corresponding to $g_{*,\chi dec}=106.75$) down to $10$~MeV (corresponding to $g_{*,\chi dec}=10.76$~\cite{Husdal:2016haj}). It can be seen that, if the light species $\chi$ makes up all the DM relic density, generating a shift of $\Delta N_{\nu}=0.1$ requires a DM mass at $m_\chi \simeq  \mathcal{O}(50)$~eV, while a shift of $\Delta N_{\nu}=0.01$ suggests that $m_\chi \lesssim 0.6$~keV. In addition, as visualized already by Eq.~\eqref{NOratio}, the shift $\Delta N_{\nu}$ would become smaller when the DM mass increases. For these sub-keV single DM candidates, they are expected to play the role of warm DM~\cite{Adhikari:2016bei}. In this case, it should be mentioned that, although the mass bounds from the Lyman-$\alpha$ forest cannot be directly applied to the non-thermal warm DM production scenarios, these candidates are generically expected to have a lower mass bound of  $\mathcal{O}(1)$~keV~\cite{Narayanan:2000tp,Viel:2005qj,Viel:2013fqw,Baur:2015jsy,Irsic:2017ixq,Palanque-Delabrouille:2019iyz,Ballesteros:2020adh}. This implies that, for a single warm DM produced via the thermal freeze-in two-body decay, the contribution to $\Delta N_{\nu}$ is negligible and hence can be probed neither by the current CMB+BBN+$Y_p$+D nor the future BBN+CMB-S4 sensitivity, no matter what the feeble coupling strength and the decaying particle mass ($m_\phi\gg 1$~MeV) are.

%------------------------------------------------------------------------------------------------------------------------------------
%---------------------------------------------------------------------------------------------------

Although the current and future BBN precision cannot be applied as a sensitive avenue to constrain or test the scenario where the light species $\chi$ makes up all the DM relic density, the future BBN+CMB-S4 sensitivity could be powerful in testing the mixed cold-plus-warm DM scenarios~\cite{Palazzo:2007gz,Boyarsky:2008xj,Diamanti:2017xfo,Murgia:2017lwo,Gariazzo:2017pzb}, where  the Lyman-$\alpha$ constraints become weaker if the particle $\chi$ makes up only a small portion of the current DM relic density~\cite{Boyarsky:2008xj,Diamanti:2017xfo,Gariazzo:2017pzb}. In this context, the simple scaling of the ratio $R_\chi$ can serve to correlate the information of the warm DM component with the observation of $\Delta N_{\nu}$ excess in the future. As an example, we show in Fig.~\ref{Rmchi-contour} the case where $\omega_\chi =20\%\,\omega_{\rm DM}$. It can be seen that the future sensitivity with $\Delta N_{\nu}=0.01$ can probe the sub-component DM at level of $\mathcal{O}(0.1)$~keV. Note that such a non-thermal warm DM component cannot be simply ruled out by the cosmological constraints discussed in Refs.~\cite{Boyarsky:2008xj,Diamanti:2017xfo,Gariazzo:2017pzb}. The reason is that the warm DM component was presumed therein to have a thermal or a thermal-like momentum distribution, while for the non-thermal DM considered here, the momentum distribution $f_\chi(p)$ is a non-trivial function of momentum and it also depends on the unknown mass spectrum $m_{\phi,\psi}$. As a consequence, there is no definite lower mass bound on the non-thermal sub-component DM due to its dependence on both $\omega_\chi$ and $f_{\chi}(p)$. Further comprehensive numerical analyses in light of astrophysical and cosmological observables are beyond the scope of this work and will not be concerned here. Rather, we would like to highlight a simple application with respect to the ratio $R_\chi$: once a small warm portion of the DM relic density is observed, e.g., via its effect on the small-scale structures, together with a possible $\Delta N_{\nu}$ excess, we can then use the ratio $R_\chi$ as an economic criterion to infer whether the origin of these two phenomena comes from a sub-keV warm DM component generated by the thermal freeze-in two-body decay.
 
\section{Validity and application of \boldmath{$R_\chi$}} 

The simple scaling of the ratio $R_\chi$ given by Eq.~\eqref{NOratio} is a common feature expected for the thermal freeze-in two-body decays, as well as the general situations where the coupling constant governing the DM production can be factored out from the Boltzmann integrals and the mass spectrum satisfies the hierarchy $m_\phi\gg m_{\psi,\chi}$. If the DM production is dominated by a single channel, we can generally factor out the feeble coupling so that the ratio $R_\chi$ is independent of this information, while the extraction of the decaying mass is possible only when it is the primary scale involved in the production channel. If $m_\psi$ is comparable to $m_\phi$, on the other hand, there would be an additional scale-dependence arising from the integration of the energy (number) yield $Y_{\chi,\rho(n)}$, and the ratio $R_\chi$ would subsequently depend on the measure of the phase-space closure in the decay:
\begin{align}\label{R-mass}
  R_\chi\propto\frac{Y_{\chi,\rho}}{Y_{\chi,n}}\propto\frac{(1-m_\psi^2/m_\phi^2)^2}{1-m_\psi^2/m_\phi^2} =1-m_\psi^2/m_\phi^2. 
\end{align}
Noticeably, this additional mass spectrum also appears in the first~\cite{Heeck:2017xbu} and the second moment~\cite{Kamada:2019kpe} of the DM momentum distribution $f(p)$, which are defined, respectively, by
\begin{align}
	\langle p \rangle &=\frac{\int d^3 p ~p f(p)}{\int d^3 p  f(p)}\propto 1-m_\psi^2/m_\phi^2,\\[0.15cm]
	\sigma^2 &=\frac{\int d^3 p ~p^2 f(p)}{\int d^3 p  f(p)}\propto (1-m_\psi^2/m_\phi^2)^2.
\end{align}
Furthermore, it has been found that decreasing both of these two moments via a quasi-degenerate mass spectrum $m_\phi\approx m_\psi$ can dilute the Lyman-$\alpha$ constraints~\cite{Heeck:2017xbu,Kamada:2019kpe}. Intuitively, this corresponds to transferring the energy of the decaying particle $\phi$ mostly to that of $\psi$ so that the DM momentum inherited from the two-body decay is suppressed, making therefore the keV warm DM colder. Nevertheless, the phase-space suppression in this case also reduces the ratio $R_\chi$, as can be seen from Eq.~\eqref{R-mass}. Thus, to have a significant shift of $\Delta N_{\nu}$ by a single sub-keV DM, we cannot simply evade the Lyman-$\alpha$ constraints with a quasi-degenerate spectrum $m_\phi\approx m_\psi$, since the contribution to $\Delta N_{\nu}$ is simultaneously suppressed. In this way, we confirm therefore the conclusion made already in Refs.~\cite{Merle:2015oja,Ballesteros:2020adh} that a single warm DM candidate under the Lyman-$\alpha$ constraints cannot generate an observable shift of $\Delta N_{\nu}$. 

It should be emphasized again that the application of the Lyman-$\alpha$ constraints discussed in Refs.~\cite{Merle:2015oja,Ballesteros:2020adh} necessitates an explicit determination of the DM momentum distribution $f(p)$. Certainly, the calculation of $\Delta N_{\nu}$ would be straightforward once $f(p)$ is solved from the unintegrated kinetic equations. In the simple algorithm presented here, however, we start from the integrated Boltzmann evolution. As shown above, this allows a fast and intuitive estimate of $\Delta N_{\nu}$ in terms of the DM information itself (i.e., the DM mass and relic density), but without a sensitive dependence on the unknown production scale and coupling. Furthermore, such a \textit{freeze-in independence} property could also be applied as an economic criterion to check,  whether a small portion of the DM relic density from a sub-keV warm component is responsible for a possible $\Delta N_{\nu}$ excess in the future.

We would like to mention that, if the DM has a thermal history, the relic particle-number and energy densities can be easily predicted with a thermal distribution. It is then straightforward to obtain that the ratio $R_{\chi}$ in this case has a similar form to Eq.~\eqref{NOratio} up to a small numeric correction. For generating an observable shift of $\Delta N_{\nu}$ from a light thermal DM, the Lyman-$\alpha$ constraints will directly come into play. Consequently, the single thermal warm DM is readily ruled out, while the thermal component in the cold-plus-warm DM scenarios would face a strong constraint from the Lyman-$\alpha$ forest data.

As a final remark, we should emphasize that the \textit{freeze-in independence} is a result of the thermal freeze-in mechanism, where the scalar $\phi$ can decay into and scatter with the SM particles in realizing the $\phi$ thermalization in the early Universe. Given a great deal of DM scenarios, it is not true that the simple scaling $R_\chi$ specified by Eq.~\eqref{NOratio} always exists. For instance, if $\phi$ is also a component of the hidden sector where some dark $Z_2$ symmetry resides at the $\phi-\chi$ sector, the super-WIMP mechanism~\cite{Feng:2003xh} can be at work to generate $\chi$ in late-time decay after $\phi$ freezes out from the SM plasma. In this case, $\phi$ is a WIMP while $\chi$ the super-WIMP DM, and one will then get the following scaling:
\begin{align}
 Y_{\chi,n}=Y_{\phi,n}\Big|_{T_{\phi,fo}},\quad	C_{\chi,\rho}\propto  y^2\mathcal{F}[m_\phi,T,f_{\phi}(y,m_\phi)],
\end{align}
where $Y_{\chi,n}$ is the relic abundance inherited from the yield $Y_{\phi,n}$ after $\phi$ has decayed away to $\chi$. $Y_{\phi,n}$ is determined at the frozen-out temperature $T_{\phi, fo}$, which depends on the particle information of $\phi$, including its gauge representation, interactions and mass. $y^2$ is extracted from the amplitude squared in Eq.~\eqref{C-rho}, while $\mathcal{F}$ is a calculable function in which $f_{\phi}(y,m_\phi)$ is no longer a thermal distribution function due to the late-time out-of-equilibrium decay and hence depends on the decay coupling $y$ and the decaying scalar mass $m_\phi$. This indicates that the ratio $Y_{\chi,\rho}/Y_{\chi,n}$ will depend on $(y,m_\phi)$ in a non-trivial way. Therefore, if $\phi$ and $\chi$ form a hidden dark sector such that the super-WIMP mechanism plays the role for $\chi$ production, the simple scaling of $R_\chi$ in terms of the DM mass cannot be realized. 

\section{Conclusion} 

If the hidden light particle has a relativistic history at the BBN epoch while becomes non-relativistic prior to the CMB regime, it is in principle possible to probe these light species via the BBN precision measurements. Motivated by this, we have demonstrated in this work that, in the thermal freeze-in production regime, the current and future BBN sensitivities cannot probe a single keV-scale warm DM under the severe constraint from  Lyman-$\alpha$ forest data, since the latter renders only a negligible contribution to $\Delta N_{\nu}$. However, the future forecast from CMB-S4, when combined with the BBN measurements, has the potential to test the warm component of cold-plus-warm DM scenarios. Such a simple observation presented here depends only on the particle-number and the energy evolution, the ratio of which can be applied to predict the shift $\Delta N_{\nu}$, basically without the sensitive information of the DM freeze-in scale and feeble coupling. 

While the simple scaling of $R_\chi$ given by Eq.~\eqref{NOratio} exists in the standard freeze-in mechanism, it is not always true for some other interesting DM scenarios. Nevertheless, we expect that this project deserves further studies in the future, as it brings us an interesting and simple avenue to probe the light hidden species that are otherwise difficult to probe directly at terrestrial colliders and indirectly in flavor physics.

\section*{Acknowledgments}

We thank Xun-Jie Xu for hospitable discussions at the early stage. This work is supported by the National Natural Science Foundation of China under Grant Nos.~12075097, 12047527, 11675061 and 11775092, as well as by the Fundamental Research Funds for the Central Universities under Grant Nos.~CCNU20TS007 and 2020YBZZ074.

\bibliographystyle{apsrev4-1}
\bibliography{reference}

\end{document}